# A Fourier Transform Approach for Automatic Detection of Oysters' Spawning

Augustine Ukpebor, James C. Addy, Kamal Ali, and Ali A. Humos

*Abstract*—Various studies have been developed to monitor the gaping behavior of bivalves (oysters) in response to environmental factors. This work aims to fully automate oyster spawning detection in real-time by building on previous efforts. The sensor system developed at Jackson State University, Mississippi, employs the Hall effect phenomenon to accurately measure the gaping of bivalves accurately. The system uses a Hall effect sensor and a small magnet glued outside the bivalve shells. The Hall effect sensor reports the magnet distance, and hence the gape at a rate of 10 Hz, which the user may change. The data generated is stored on an SD card and supplied to a microcontroller for transmission to our file server on the Internet. The collected time series data is transformed to the frequency domain in 10-minute chunks using the fast Fourier transform (FFT) algorithm. When the data is analyzed in the frequency domain, the spectral power in the frequency range of 0.3 Hz to 1.3 Hz spikes several orders of magnitude when spawning occurs. It is then concluded that using a threshold of 0.1 dB in the frequency range of 0.3 Hz to 1.3 Hz, and spawning could be predicted 100% of the time.

## I. INTRODUCTION

1.1 Overview

In the context of the emerging Internet of Things (IoT), the advances in embedded systems and sensor technologies make it possible to connect specific devices to the Internet. This study describes the experimental field and laboratory trials designed to use sensor systems to digitally collect data from bivalves and transmit it to a file server on the Internet. The system employs the Hall-effect phenomenon to accurately measure and report the gaping of oysters [1]. The research deployment uses a Hall-effect sensor and small magnet, which are glued to the exterior of the shells of an oyster. A Hall effect sensor is a universal magnetic field sensor with a linear output. The sensor measures the distance of the magnet to a microcontroller that records and sends the data to a base station for real-time analysis [2].

In general, to address the inefficiency of visual inspection of spawning events by experts, we analyze the collected data from oysters to evaluate certain behaviors called spawning.

This paper was submitted on September 4, 2022. This project was paid for with federal funding from the Mississippi Department of Environmental Quality and the Department of the Treasury under the Resources and Ecosystems Sustainability, Tourist Opportunities, and Revived Economies of the Gulf Coast States Act of 2012 (RESTORE Act). The statements, findings, conclusions, and recommendations are those of the author(s). They do not necessarily reflect the views of the Mississippi Department of Environmental Quality or the Department of the Treasury. Thanks to Dr. Isaac Osahon, Chao Jiang, Paul C. Ekeziel, and Chinonso Ezeobi for the paper review.

Augustine Ukpebor, James C. Addy, Kamal Ali, and Ali A. Humos Jackson State University, Jackson, MS 39217, USA.
augustine.ukpebor@students.jsums.edu,
james.c.addy@students.jsums.edu, kamal.ali@jsums.edu, and ali.a.humos@jsums.edu.

Spawning is the process of releasing eggs and sperm, and it occurs during reproduction. More precisely, we demonstrate how the signal level is dispersed over the frequency domain using the fundamental mathematical techniques of FFT and power spectral density. We generated data from two sets of six oysters with gaping activities measured over 90 days. The information and associated statistical analysis are displayed graphically and made accessible to the public.

1.2 Related Work

This section reviews previous relevant research on three aspects: This first explains the precise timing of spawning in oysters. The second reviews simple approaches to online data collection. Finally, an algorithm to determine the spawning of oysters from the velocity signal is examined.

The interest in studying the activities of bivalves by measuring their motions is not new [3]. Marceau [4] was a forerunner in this field, as he was the first to record mollusk valve motions [3], [4]. In recent years, there has been clear research interest in directly measuring the behavior of mussels under natural conditions [3], [5], [6], [7]. A few years ago, a method for measuring the shell gape in bivalves was introduced by [8] and adapted to a scaled-down version of equipment used with marine endotherms [9]. Utilizing the Hall effect, the method consists of a Hall effect sensor, a magnetic sensor that can convert magnetic fields into a measurable voltage output. The strength of the magnetic field passing via the sensor is directly proportional to the output voltage. It can, therefore, continuously record the movement of a magnet relative to the sensor. By attaching a magnet and Hall sensor to either valve, the shell gape of bivalves can be observed continuously without intrusion.

1.2.1 Timing of Spawning in Oysters

The exact timing of spawning is vital from a demographic point of view for marine animals living in high-energy coastal habitats [10], [11]. This critical timing is especially proper for broadcast spawners because males and females must discharge their gametes into the open sea. The hydrodynamic process results in optimal encounter rates for zygote mixing, fertilization, and dispersion. Oyster farmers understand that rapid temperature pulses of 5-10 ° C cause the spawning of mature oysters [11], [12]. The articles [11], [13], [14] reported that raising the water temperature can encourage the spawning of male and female oysters. Galtsoff [13], [14], and Bernard [11] also discovered that the presence of sperm could cause females to spawn. Korringa [15] indicated that Crassostrea edulis most frequently spawned during spring tides, at two days after full and new moons. Nelson [16] and Bernard [11] reported that spawning takes place during or just following the flood tide, and Prytherch [17] and Bernard [11] noted that this incident happened during spring tides close to or at high water (HW).

1.2.2 Online Data Collection

There is a compelling need to employ remote Internet sensors to communicate information promptly and broadly every day, particularly in marine environments, due to increased interest in the health of aquatic systems [18]. In one of the monitoring systems developed recently, sensors are configured to read gaping data and transfer it to a repository server in the cloud [2]. A system is built with a platform capable of transmitting gaping data continuously from the field to the Internet, where an artificially intelligent system analyzes the data in real-time [1].

1.2.3 Spawning Detection - Monitoring the Behavior of the Bivalves

An algorithm is proposed to determine spawning from the velocity signal based on the fault detection approach. This technique works by estimating the rate of movement of the valve by calculating the time derivative of the valve distance [19]. Another method of assessing how bivalves behave in response to environmental exposure is observing their opening and closing behaviors. One potential means of assessing the water quality is through the ability of mollusk bivalves to taste their surroundings. Bivalves open and close their valves in a defensive reaction to external stimuli such as touching or shading, or by the sudden approach of a predator, as well as in response to a deteriorating environment due to toxic red tides, oxygen deficit, or low salt concentrations [20], [21]. Valve movements of bivalves are closely related to physiological processes such as respiration, nutrition, reproduction, and excretion, which are modulated by environmental parameters [6], [22]. Furthermore, an accurate understanding of the gaping of bivalves plays a vital role in managing water quality in shellfish habitats as well as the health of the shellfish.

1.2.4 Contributions

Our proposed algorithm for detecting oysters' spawning differs from previous studies by substantially addressing the prior gaps. One of the gaps this study addresses is the fault detection approach. It works through the estimation of velocity by selecting different algorithm parameters by trial and error [16]. The following lists novel contributions of this work:

1. An efficient algorithm is explored for the automatic detection of oysters' spawning. This technique revolutionizes the manual approach used by aquaculture experts to manage oyster behavior which sometimes can be inaccurate. This approach is proven to be highly precise and stable.
2. We developed a cost-effective and highly efficient gape measurement system to monitor the behavior of oysters in real-time.

II. RESEARCH METHODOLOGY

2.1 Research Location

The in-situ experiments occurred at three locations in Mississippi, United States: The JSU Research Laboratory, Jackson, The University of Southern Mississippi Gulf Coast Research Laboratories (GCRL), Ocean Springs, and Mississippi Department of Marine Resources (MDMR), Pass Christian. The JSU research lab is where the sensors are built and tested before deployment, while GCRL and MDMR are the experimental locations. The GCRL is a marine research center and is a unit of The University of Southern Mississippi's College of Science and Technology. The GCRL operates national research and development programs in fisheries, geospatial technologies, marine aquaculture, environmental assessment, and marine toxicology. Geographically, the city of Pass Christian is located on the Mississippi Sound and situated on a peninsula. The city is surrounded by water to a long stretch of the bayou to the north. The Gulf of Mexico bounds it to the south and the Bay of St. Louis to the west.

2.1 Experimental Setup

This research's electronics and mechanical hardware components were fabricated to suit project objectives. The sensor components were chosen because of their unique functionalities and low cost. The sensor system comprises the Hall effect sensor (HAL 2425), microcontrollers (Arduino nano and Wemos ESP8266/ESP32), and magnets. Other hardware components employed are Secure Digital (SD) card reader, calibration tool, soldering materials, digital meter, cables, Polyvinyl Chloride (PVC) pipes, couplers, flex seal, and a 12V DC Battery. A Uprint 3D printing machine was used to produce various project components to house the microcontrollers.

Fig. 2.1 represents the step-by-step research process. First, it involves gathering the research materials as discussed in the preceding paragraph. Next is the calibration and linearization of the Hall effect sensors, alongside the design and printing of sensor holders. Then comes the provisioning of sensors, followed by the building and testing of codes. After successfully testing the sensors, the codes are loaded into the Arduino nano and Wemos and they are tested afterward. Finally, we deploy the sensors, and then transmit data through the system.

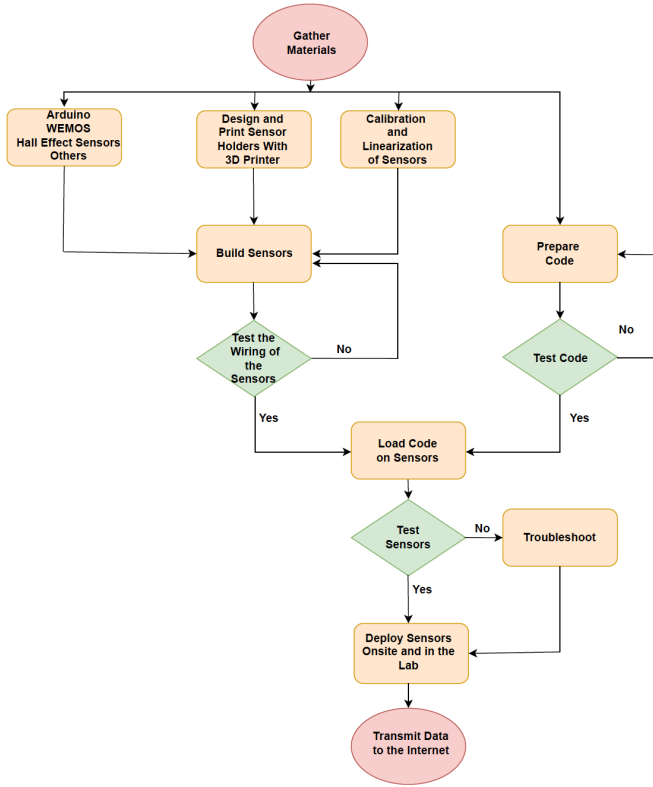

**Fig. 2.1.** Research step-by-step process.

2.2 Design and Overview of The System

The system's architecture, discussed in our previous study [2], comprises three units: one microcontroller unit, an Arduino nano, another microcontroller unit, the Wemos - ESP8266, and six Hall effect sensors, HAL 2425. For the rest of this paper, Hall effect sensors, HAL 2425, Arduino nano, and Wemos - ESP8266 will be referred to as Hall effect sensors, nano, and Wemos respectively. Fig. 2.2 shows the block diagram of the sensor system.

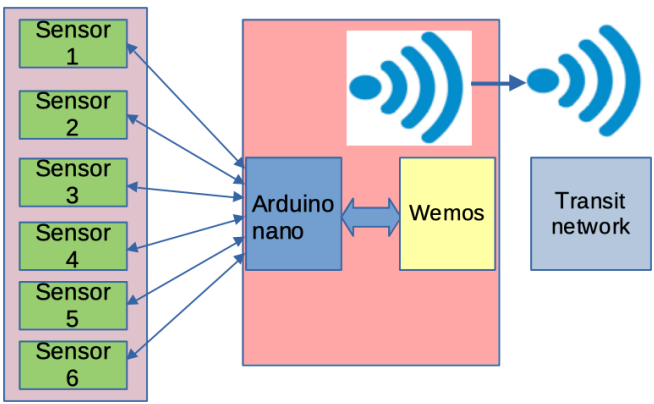

**Fig. 2.2.** Block diagram of the sensor system.

The Hall effect sensor and Wemos are connected to the nano. The Hall effect sensor measures the valve gaping of the oysters. The nano reads data from the sensors and stores them in an array. The Wemos requests data from the nano at a predetermined interval and sends it through the transit network to the external TELCO Base Station.

2.3 Data Collection and Data Transmission

The research data was collected from six oysters and transmitted to a remote file server on the Internet. The remote server is provided using a hosting web service by Hostinger. The data is available to the public via a research website. In brief, the data collection process is easy. The algorithm that runs on the Micro Controller Unit (MCU) has two variants: one with a temperature sensor and the other without a temperature sensor. This research uses the former. The Hall effect sensors measure the gape activity of the oysters. Then, the nano reads six sensors and stores six bytes of data in an array. Next, the Wemos requests data from the nano through the Inter-Integrated Circuit (I2C), and the nano sends back the six bytes to the Wemos. The I2C protocol is a serial communication protocol used to connect low-speed devices and integrated circuits (ICs), in this case, the nano and the Wemos.

2.4 Procedure of Data Analysis

Time and frequency domain techniques were used to analyze the data captured from the oysters. The time domain was used to evaluate the data to visualize the animal behavior casually. After the data preprocessing, MATLAB was used to plot the data, and the axes were properly formatted to scale and represented with the appropriate units. Given that visual inspection of time domain graphs comes with inherent human errors, FFT was used to provide more insight into the frequency components hidden deep within the data. We could determine the frequencies being excited and the amplitude at each frequency.

Additionally, we could highlight changes in the frequency and amplitude and the harmonic excitation in the broad frequency range. Lastly, we computed the average power to better understand the signal's frequency components. The average power was used to determine the spawning events of the oysters. The experiments and results section substantially presents the details of the analyzed data.

### III. RESULTS AND DISCUSSION

3.1 Dataset

The dataset used in this research is from a repository file server located on the research website. In our case, the dataset for 2018 collected from oysters deployed in the field at the GCRL has been used. Dataset 1 (Table I) is a one-week dataset from September 3 - 10, 2018, excluding September 9, 2018 (insufficient data) because of the Wi-Fi connection challenge. The data is a time series dataset comprising six features and 398,404 records. Similarly, dataset 2 (Table II) is a week of data from October 25 – 31, 2018. Dataset 2 is a time series dataset with six features and 425,233 records.

## TABLE I
### DATASET 1 (SPAWNING)

| S/N | Date | Data readings are taken ten times per second | Data readings (seconds) | Hours |
|---|---|---|---|---|
| 1 | 3-Sep-18 | 231430 | 23143 | 06.40 |
| 2 | 4-Sep-18 | 861300 | 86130 | 23.91 |
| 3 | 5-Sep-18 | 861830 | 86183 | 23.90 |
| 4 | 6-Sep-18 | 785240 | 78524 | 21.80 |
| 5 | 7-Sep-18 | 470440 | 47044 | 13.10 |
| 6 | 8-Sep-18 | 110060 | 11006 | 03.10 |
| 7 | 10-Sep-18 | 663740 | 66374 | 18.40 |
| TOTAL | | 3984040 | 398404 | 110.61 |

## TABLE II
### DATASET 2 (NON-SPAWNING)

| S/N | Date | Data readings are taken ten times per second | Data readings (seconds) | Hours |
|---|---|---|---|---|
| 1 | 26-Oct-18 | 140090 | 14009 | 03.89 |
| 2 | 27-Oct-18 | 859520 | 85952 | 23.88 |
| 3 | 28-Oct-18 | 848670 | 84867 | 23.57 |
| 4 | 29-Oct-18 | 862670 | 86267 | 23.96 |
| 5 | 30-Oct-18 | 862100 | 86210 | 23.95 |
| 6 | 31-Oct-18 | 679280 | 67928 | 18.87 |
| TOTAL | | 4252330 | 425233 | 118.12 |

3.2 Data Preprocessing

Data preprocessing is an essential technique in data mining as it goes through a series of steps such as data cleaning, data integration, transformation, and reduction. In our case, only data cleaning was performed. To clean up the data, MATLAB was used to remove inconsistencies; the inconsistencies include the same data in different formats, missing data, and empty rows. As a preprocessing step, we removed empty rows with missing data, concatenated the available data, then ignored the parts where data did not exist. Next, we smoothed possible noise data to obtain good frequency resolution, took an average of every ten values in the array, and normalized the dataset to have a zero-starting point. The sensor system takes readings ten times per second from the oysters. Hence, we took the mean of every ten records to realize a data reading of once per second.

3.3 Spawning Graph from Time Domain Analysis

Fig. 3.1 is a subplot of data from oysters 2, 3, and 6. The data has 398,404 samples. The vertical axis represents the valve gape amplitude in millimeters, while the horizontal axis represents time in hours. Given the three graphs, there is a significant difference in the valve gape amplitude as exhibited by the three oysters, with the gaping activities more prominent in oyster 2. The opening amplitudes that were most frequently observed were in the range of 0.15 – 0.25 mm. The three oysters have maximum amplitude of 0.30 mm. The valve gape is the distance between a valve pair which translates to the opening and closing of the shells.

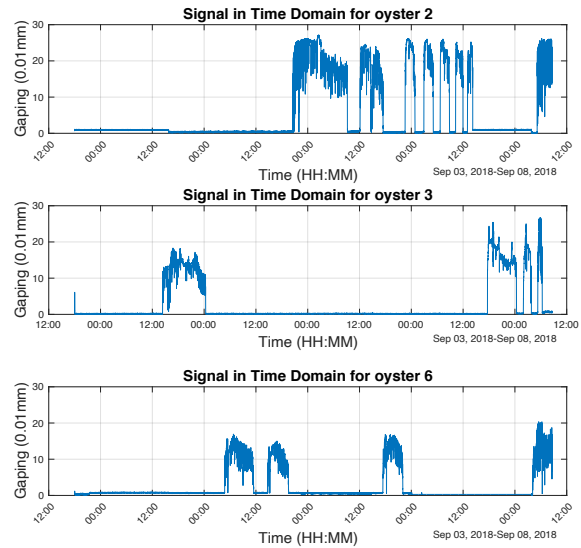

**Fig. 3.1.** Spawning result for dataset 1 with date and time.

Fig. 3.1 shows that oysters were deployed at 18:00 CST. The figure also shows that the gaping behaviors happened on the second and last days for oysters 3 and 6. Spawning also occurred on the fourth day for oyster 6. It is interesting to note that oyster 2 exhibited a unique behavior from the other oysters. There was no activity (oyster 2) for the first two days of system deployment until the third day. What is intriguing about this process is that spawning events happened on the last day for the three oysters. Based on empirical evidence from our experiment, spawning could happen any time of the day; however, in this instance, spawning occurred in the morning.

Let us briefly explain why the graph is not fully represented with the required number of days, even though the data analysis is predicated on seven days. Different amounts of data were transmitted daily, and while some days had a total of 24 hours' worth of data transmitted, others had as little as only three or four hours (see Tables I and II). The fluctuation in Wi-Fi connectivity at the research center could have been responsible for the anomaly, resulting in missing or incomplete data. To remedy this inconsistency, we have concatenated the available data, then ignored parts where data did not exist. Finally, though the data had been collected over seven days, the available records were less than seven days. Finally, Fig. 3.1 shows that the analyzed data was slightly longer than four days. The overall correlated data depend on the reliability of Wi-Fi.

3.4 Plotting Frequency Spectrum Using MATLAB

This section presents how to use the FFT function in MATLAB to analyze a discrete signal to determine the frequency components of the signal (amplitudes, frequencies, and phases present in a signal).

We use a sampling frequency of 10 Hz. The data used in this research has 398,404 samples with six columns, which essentially translates to six oysters. For simplicity, we would analyze the data for one oyster by using a well-characterized method for analyzing the signal using the Fourier transform. First, the sampled data is read into an array. A moving average (MA) is used to smooth trends by filtering out the "noise" from the original signal. Next, we define samples per unit of time or space, signal length (number of samples), and space range for data/time vectors. Then, a specific sensor data array is selected, and the discrete Fourier transform (DFT) data is computed using a computationally efficient FFT algorithm. After that, we computed the two-sided spectrum. Next, the single-sided spectrum, is derived from the double-sided spectrum. Before the computation, the frequency domain, (f) is adequately defined. We use longer signals since it tends to produce better frequency approximations. Finally, the MATLAB "stem" function is used to plot the single-sided spectrum at values specified by frequency.

The horizontal axis of Fig. 3.2 is the normalized frequency where the frequencies are all normalized to the Nyquist frequency, which is half the sampling frequency.

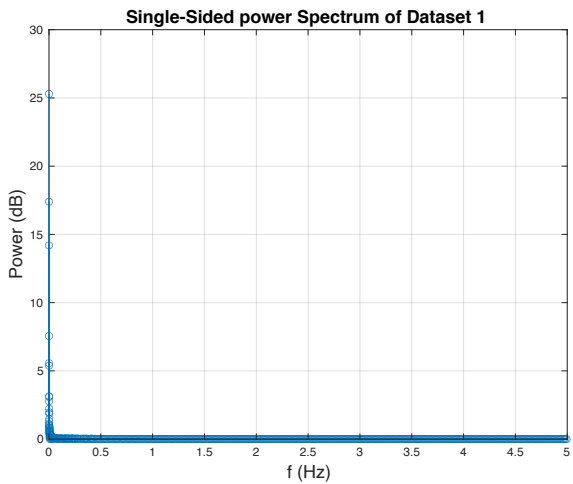

**Fig. 3.2.** Single-sided power spectrum for spawning event (dataset 1).

Given the graph, the maximum power is 25.1 dB at a frequency of about 0.01 Hz. The exciting fact about frequency spectrum analysis is that it provides detailed signal characteristics, including the frequency components. It also shows how much of the frequencies are concentrated over a range of frequencies. The graphs (see Figs. 3.2 and 3.3) of the frequency spectrum of datasets 1 and 2 looks similar as further analysis reveals otherwise. The zoom-in of the graph provides how much signal is present within a given frequency band. We will discuss this process extensively in the subsequent sections.

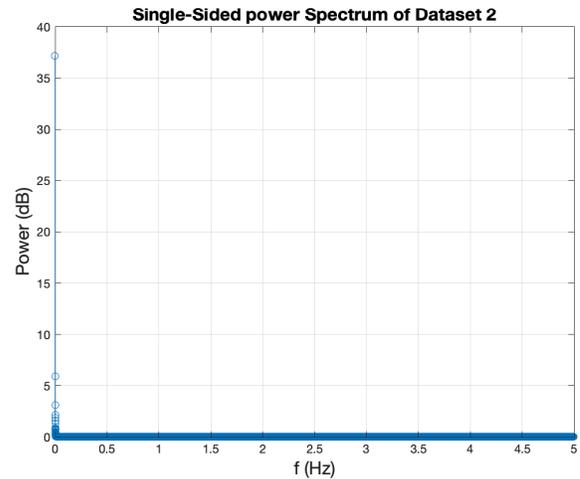

**Fig. 3.3.** Single-sided power spectrum for non-spawning events (dataset 2).

3.5 Spawning Events

Further analysis of Fig. 3.2 produces Fig. 3.4, which is the zoom-in of the selected signals, 1200 readings (20 minutes). The portion of the selected signal shows spawning events. The interesting observation is that the power distribution is heavily concentrated in the region of the spawning events.

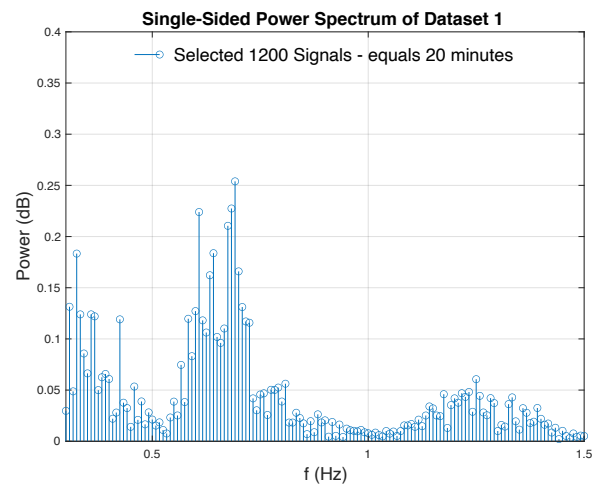

**Fig. 3.4.** Zoom-in spawning for 1200 readings - 20 minutes for dataset 1.

## 3.6 Non-Spawning Events

Similarly, further evaluation of Fig. 3.3 produces Fig. 3.5, which is the zoom-in of the selected samples, 1200 readings. Now, the portion of the selected signal shows non-spawning events. In contrast, the power distribution within the zoom-in region is negligible as against the case of the spawning events.

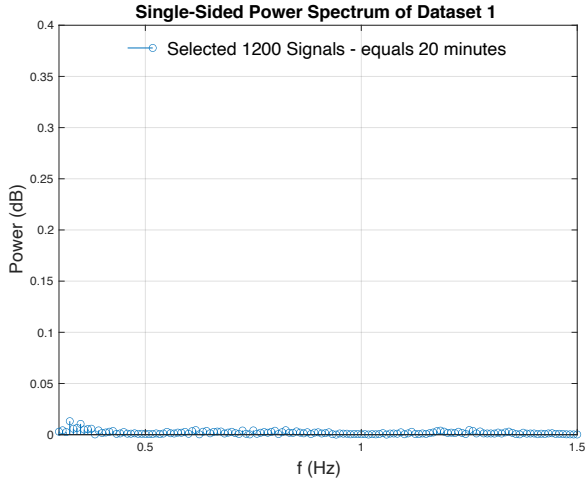

**Fig. 3.5.** Zoom-in non-spawning for 1200 readings - 20 minutes for dataset 2.

## 3.7 Spawning Graph in Time and Frequency Domain

This section explains the spawning activities of the oyster for both time and frequency domains. We take readings every 0.1s successively (10 Hz) for every oyster. With a sampling frequency of 10 Hz, we use FFT to analyze the frequency response of data by segmenting the Fourier transform as a function of frequency to visualize how the frequency content changes over time. The overall goal is to see sharp spikes in frequency. First, we take the FFT of a particular data of an oyster. Next, the FFT of the last n rows (in this case, 6,000) from the bottom is taken. Next, we repeat the previous step for the next 6,000 rows until it gets to the first row or data is exhausted. Lastly, we combine the time column with any column of the data to plot the graph in both the time and frequency domains (see Fig. 3.6).

Fig. 3.6 is the graphical representation of the three subplots. The top and middle subplots are the zoom-in frequency and time domains for selected signals (6,000 samples equivalent to 100 minutes), respectively, while the bottom subplot is the time domain for the entire signal of 18 hours and 21 minutes. At first glance, one can see that some patterns emerge on the frequency spectrum. For example, there appears to be a strong signal for the first few thousand samples with more complicated details, but we cannot easily say much about this signal. However, if we could reduce these data to a manageable size, more revelations could be revealed (see Fig. 3.4).

The spawning portion of the bottom graph (Fig. 3.6) is zoomed-in to produce the middle graph to reveal a complete detail about the frequency components could be revealed. Lastly, we examined only the signal with spawning activities for the time domain signal (middle subplot). The spawning happened for about one hour and forty minutes starting at 21:06 and ending at 22:46 (see the middle subplot of Fig. 3.6).

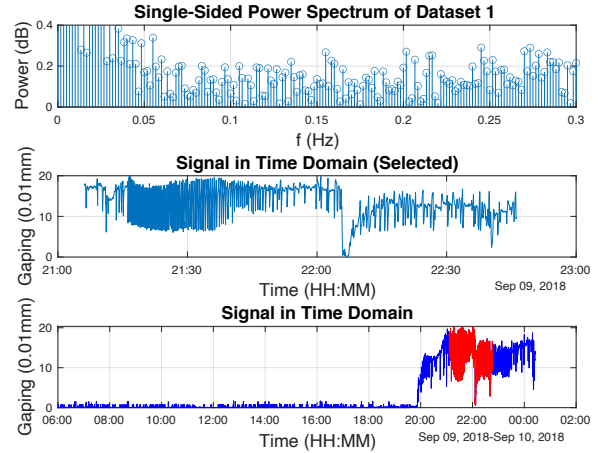

**Fig. 3.6.** Zoom-in time and frequency domain for dataset 1.

The two uppermost subplots of Fig. 3.7 are the zoomed-in portions of the frequency and time domain signals, respectively. The power spectrum evaluation (top subplot) shows no spawning behavior since there is no spike frequency in the entire portion of the frequency spectrum.

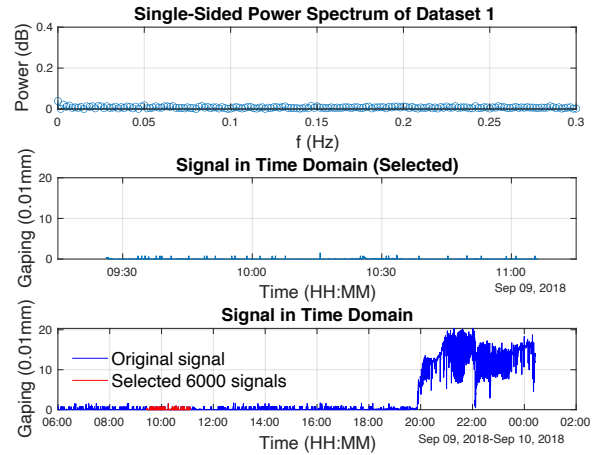

**Fig. 3.7.** Zoom-in indicating non-spawning events.

## 3.8 Discussion

The results from this section show the average power (dB) from the experiment described in sections 3.4 and 3.7. To evaluate the spawning behavior of oysters, we present the average power results for various data samples of 100, 300, 500, 1000, 2000, and 6000 from a large dataset. The data from Table I was used for the simulation using MATLAB; the data has 398,404 samples with six columns, which essentially translates to six oysters. However, for this implementation, we use the data from oyster 6 for the analysis. It should be noted that data samples of less than 100 points are not considered in this research because of their computational complexity. Upon graphing the data from oyster six and subsequently zooming in, we observe the power spectrum is heavily concentrated at lower frequency ranges between 0.3 Hz and 1.3 Hz, with little or no power outside the range. Again, significant spikes in frequency suggest a possible spawning area exists.

Further evaluation of frequency components shows some patterns, signifying potential erratic behavior at an average power greater than 0.1 dB. When the average power of samples of 6,000 records was calculated, we observed only two points greater than or equal to 0.1 dB. Additionally, computing 2,000 and 1,000 points, we recorded 6 and 12 points, respectively, greater than or equal to 0.1 dB. Again, evaluating 500 and 300 samples produced 24 and 39 points correspondingly greater than or equal to 0.1 dB. Finally, 100 samples produced 365 points. The last result explains why it is computationally intensive to compute the average power for significant data points while analyzing 100 records simultaneously. Based on the previous analysis, the algorithm accuracy is higher when few signal samples are considered; however, with a computational or overhead trade-off. Consequently, the magnitude of the average power depends on a few factors such as sampling frequency, number of readings, and frequency range. This is the reason the value varies slightly for the six cases. Generally, the graphs of the average power data for the six records are shown in Fig. 3.8.

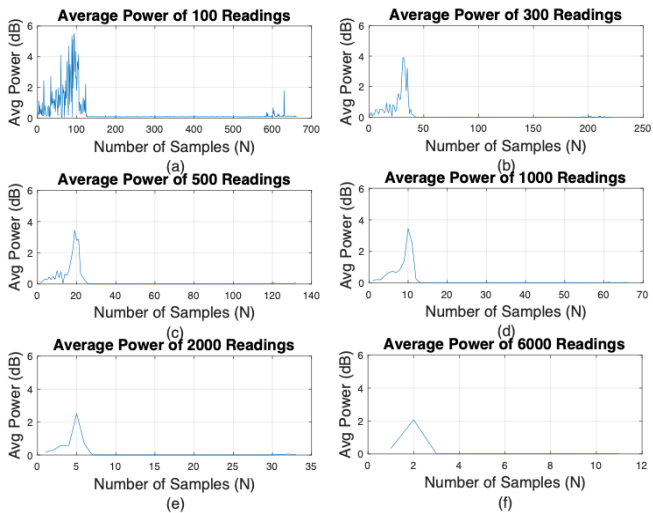

**Fig. 3.8.** Average Signal Power for six data samples. (a) 100 samples. (b) 300 samples. (c) 500 samples. (d) 1000 samples. (e) 2000 samples. (d) 6000 samples.

Finally, we have seen how to change the frequency range to obtain good frequency resolution and frequency components of interest in this case, with the minimal noise level, and we obtain good spectral details. Therefore, given our investigation of the previously discussed 6000 records (average power), any power of 0.1 dB or higher is spawning behavior, and power less than 0.1 dB are NO spawning events. This study demonstrates that the 0.1 dB threshold works for other data as illustrated in Fig. 3.8. Our investigation also shows that the graph gets smoother as the number of samples increases, as elucidated in Fig. 3.8. Optimal accuracy is achieved with fewer signals, though with a trade-off of overhead.

## IV. CONCLUSIONS

In this work, we present an efficient algorithm for automatically detecting oysters' spawning. The proposed algorithm is based on evaluating the average power of a given input signal. Results from the experiment show that the power spectrum at lower frequency ranges between 0.3 Hz and 1.3 Hz is quite significant, with little or no power outside the range. This result means that during spawning events, the oyster opens their valve at a rate of at least 0.3 times and at most 1.3 times. Therefore, given the investigation of 6,000 data points, it is established that any power of 0.1 dB or higher signifies spawning behavior, and power less than 0.1 dB are NO spawning events. The study also demonstrates that an average power of 0.1 dB threshold works for various sample data. The algorithm was tried with all possible scenarios, and the results look promising. Additionally, the experiment shows that oyster spawning events can be automatically detected both in the laboratory and in the field when the algorithm is applied. It should be noted, however, that the accuracy seems to be impaired when computing recursive data points of less than 100 records.